\documentclass[journal]{IEEEtran}
\usepackage{placeins}
\usepackage{afterpage}
\usepackage{pdflscape} % or lscape
\usepackage{rotating}
\usepackage{}
\usepackage{pifont}
\usepackage{amssymb}
\usepackage{array}
\usepackage{ragged2e}
\newcolumntype{L}[1]{>{\RaggedRight\arraybackslash}p{#1}}
\usepackage{rotating}
\usepackage{tabularx}
\usepackage{xparse}

\usepackage[utf8]{inputenc}
\usepackage{cite}
\usepackage{graphicx}
\usepackage{amsmath,amssymb}
\usepackage{booktabs}
\usepackage{tabularx}
\usepackage{hyperref}
\hypersetup{hidelinks}
\usepackage[caption=false,font=normalsize,labelfont=sf,textfont=sf]{subfig}
\usepackage{pbalance}

\newcommand{\MonthName}{%
\ifcase\month\or January\or February\or March\or April\or May\or June\or July\or August\or September\or October\or November\or December\fi}
\newcommand{\richardsdate}{\MonthName\ \number\day, \number\year}

\title{Electrical Power Network Modeling Framework for Wildfire Risk and Resilience Analysis}

\author{Richard~Campos,~\IEEEmembership{}
        Erica~Fischer,~\IEEEmembership{}
        Eduardo~Cotilla-Sanchez,~\IEEEmembership{Senior Member,~IEEE,}% <-this % stops a space
\thanks{R. Campos is with the School of Civil and Construction Engineering, Oregon State University, Corvallis, OR, 97331 USA e-mail: camposri@oregonstate.edu.}% <-this % stops a space
\thanks{E. Fischer and E. Cotilla-Sanchez are with Oregon State University.}% <-this % stops a space
\thanks{Manuscript received \richardsdate}}
\begin{document}
\maketitle
\bstctlcite{IEEEexample:BSTcontrol}

\begin{abstract}
The increasing intensity and frequency of wildfires are causing significant economic and societal impacts on communities through direct effects on the built environment, particularly critical infrastructure.
Electrical systems can both initiate wildfires (grid-to-fire) and be damaged by wildfire exposure (fire-to-grid).
Therefore, resilient electric systems can both limit ignitions and be hardened such that they are more robust to fire demands.  
Researchers have investigated wildfire mitigation strategies using traditional transmission and distribution electrical test-system models.
However, these test cases may not accurately represent realistic electrical system configurations or fuel landscapes, nor capture community impacts, particularly the social and economic effects of mitigation strategies.
A wildfire-aware modeling framework enables researchers to develop test cases that benchmark resilience and mitigation strategies while reducing reliance on overly simplistic assumptions about wildfire effects on electrical systems and communities.
This study proposes a modeling framework for wildfire-electrical system research by analyzing recent literature and identifying key dimensions as well as gaps within these dimensions.
In particular, the framework considers how fire in the wildland–urban interface propagates in space and time, how hazard–infrastructure interactions (e.g., wind and fire) cause system- and component-level damage, and how wildfire-related power outages affect communities.

\end{abstract}


%%%%%%%%%%%%%%%%%%%%%%%%%%%%%%%%%%%%%%%

\section{Introduction}
\IEEEPARstart{T}{he} impact of wildfires on electrical systems and communities is exacerbated by fire weather conditions \cite{Su_2024_IEEE_TPS, Campos_2026_IJDRR}.
During fire weather conditions, electrical systems can both precipitate wildfires through ignition and be heavily affected by the fires themselves through direct heat damage \cite{Pianco_2025_IEEE_TPS,Nazemi_2024_IEEE_TPS}. 
Strong winds cause electrical-induced fires through wind-damaged electrical components and spread fires and firebrands (embers) over large distances. 
Regardless of the origin of the fire, once it has started, electrical systems are vulnerable to physical damage (e.g., power lines, poles, or transformers), resulting in operational disruptions and extended power outages. 
These impacts may cause system shutdowns to protect the rest of the system, prevent ignition, or result from direct damage to system components \cite{Vahedi_2025_}. 
For example, overhead conductors and supporting structures can be damaged by wildfires due to wildfire heat, direct exposure to flames, and wind-driven smoke \cite{Moreira_2024_IEEE_TPS}.
The resulting impacts on electrical systems from wildfires have significant implications for emergency response, critical infrastructure, and the response and recovery phases of communities following a wildfire disaster.
This study proposes a framework that summarizes the co-dependencies between the wildfire domain and the electrical grid domain considering the bidirectional impacts between the two: fire-to-grid (f2g), and grid-to-fire (g2f).

To either prevent electrical components from causing fires or reduce the impacts of outages to customers if wildfires occur near the network, electric utilities have adopted various mitigation methods \cite{Vazquez_2022_IEEE_TPS}. Some of these common methods include:
\begin{itemize}
  \item Public Safety Power Outages (PSPS):~Utilities can temporarily shut off power in high fire risk areas during red flag warnings. 
  This action can keep electrical equipment from creating sparks that may cause a fire, but it also disconnects power from families and businesses. PSPS addresses only g2f issues and does not protect equipment or portions of the system for f2g.
  \item Equipment Hardening:~Utilities can replace old equipment and poles with more fire-resistant materials, perform vegetation management around power lines and poles, or bury lines underground. 
  These actions make the system more resilient to fire, addressing g2f by reducing ignitable material around power lines, but also f2g by increasing the fire resistance of critical equipment thereby reducing vulnerability to individual component damage that may cause a cascading failure within the grid. 
  \item Automatic Shutoffs and Load Shedding:~Utilities can rapidly disconnect specific areas or reduce load during a fire event to reduce the power demand on certain parts of the system during dangerous weather conditions that may cause sparks (g2f) or create vulnerabilities within the system for damage (f2g). 
  Unlike for PSPS, automatic shutoffs occur locally in small portions of the network and occur based upon pre-determined conditions. 
  \item Islanding into Microgrids:~Islanding enables systems to isolate components by partitioning the electrical grid. 
  Through islanding, utilities can create microgrids when the island can survive independently of the bulk system. 
  This type of action allows utilities to conserve resources locally for critical infrastructure (e.g., hospitals) or to enable load shedding tactics such as PSPS or automatic shutoffs locally during red flag warnings or during a fire itself. 
  There are many benefits to this type of operational planning, as it enables utilities to maintain service in areas with less exposure or less fire vulnerability during fire weather events thereby enabling business continuity and allowing residents to maintain power. 
  However, not all grids are set up to operate in this manner, and the relay coordination to implement this type of operational planning requires considerable planning as well as trained personnel. 
  There are many active research questions about which operational planning effort is most effective (e.g., load shedding and islanding). Islanding addresses g2f by shutting off portions of the system that might ignite fires under extreme weather conditions. 
\end{itemize}

To evaluate the effectiveness of these methods, researchers have analyzed electrical systems through operations or planning studies, which often also use standard test systems. 
For example, wildfire-related electrical system studies evaluate mitigation strategies using standard IEEE test systems. However, these benchmarking testbeds were not designed to represent the coupled behavior of transmission (Tx) and distribution (Dx) infrastructure, WUI landscapes (e.g., fuel landscapes), fire behavior (e.g., wildfire propagation), or to evaluate community resilience metrics (e.g., economic or social consequences to system component damage, system damage, or any of the mitigation methods) \cite{Ansari_2014_ITEES, Izadi_2023_AE, Mohagheghi_2015_IJEPES, Rhodes_2021_IEEETPS, Vahedi_2025_}. 
A wildfire-aware test system is needed to allow for consistent benchmarking, reduce the reliance on case-specific wildfire assumptions, and support future research beyond initial focus areas.
This type of test system would also enable Tx overhead line failures to be extended to examine the effects of those ignitions on Dx-level resilience strategies using the same test framework.
Therefore, to quantify fire risk to and from electrical systems, a holistic test framework that incorporates wildfire propagation, Tx\&Dx, and downstream effects are necessary. 

In general, wildfire-related electrical studies are structured within nine dimensions:~operations/planning, system type, system case study testbed, hazard, component exposure, component response modeling, disaster phase, and community resilience.
Table~\ref{tab:lit_rotated} summarizes these dimensions for representative studies in the power-systems literature.
Further, a select few studies were included as examples to distinguish gaps in the wildfire-electrical literature.

Wildfire-related studies consider either the Tx or the Dx, where only a few studies couple both systems.
Incorporating both systems (Tx~\&~Dx) into one testbed would provide a deeper understanding of initiating events and propagation of cascading outages, voltage profile problems, and other protection issues during wildfires.
A combined test system would also provide insights into wildfire outbreak dynamics on electrical systems, where a distribution line causes a fire that ultimately impacts the transmission system or vice versa. 
Additionally, standard testbeds rarely represent the loss of power to critical infrastructure (e.g., hospitals, fire stations) or the effects of PSPS to community functionality in general.

As a result, this study aims to establish a framework for a Wildland-Urban Interface (WUI)-embedded electrical test system that can represent WUI landscapes, integrate fire hazard inputs, couple Tx~\&~Dx domains, and support community-relevant resilience metrics.  
The contributions of this paper are:~(1) to synthesize the state-of-the-art research on wildfire-related electrical resilience across consistent analytical dimensions and (2) to develop an electrical network power modeling framework for wildfires.  
The proposed framework emphasizes coupled wildfire-power interactions that can support planning, operations, and restoration/recovery analysis while also enabling community-centric metrics and objectives.
The remainder of the paper is organized as follows. 
Section~\ref{sec: Background} summarizes previous research using a consistent set of analysis dimensions that span decision-making, system type, hazard representation, exposure, response modeling, restoration, and social considerations.  
Section~\ref{sec: Proposed_Framework} introduces the framework and the requirements for a wildfire resilience or risk test system.  

%%%%%%%%%%%%%%%%%%%%%%%%%%%%%%%%%%%%%%%
\section{Synthesis of State-of-the-Art and Research Gaps}\label{sec: Background}
This section summarizes previous research within the field of wildfire-resilient electrical grids (Table~\ref{tab:lit_rotated}), and where the gaps in knowledge exist to enable a comprehensive community resilience evaluation of the electrical grid as it pertains to the electrical grid.

\subsection{Operations and Planning}\label{sec: OperPlann}
The studies synthesized in Table~\ref{tab:lit_rotated} use mitigation that can be generalized into three categories, operational, planning or both. 
Wildfire resilience literature in electrical systems is currently dominated by operational studies that analyze specific mitigation strategies implemented before and during wildfire events (Table~\ref{tab:lit_rotated}).
These types of operational studies commonly include distributed generation (DG), reconfiguration, and PSPS \cite{Mohagheghi_2015_IJEPES, Rhodes_2021_IEEETPS, Greenough_2025_IEEETPS, Nematshahi_2026_EPSR, Yao_2025_IJEPES}.  
For example, Dx operational mitigation has been explored on modified IEEE feeders using dispatch-oriented approaches and scheduling under uncertainty \cite{Ansari_2014_ITEES, Izadi_2023_AE}.  
Similarly, operational studies at the Tx level have examined generator dispatch and optimal flow (OPF)-based responses under wildfire-driven line constraints \cite{Choobineh_2015_FSJ, Nematshahi_2026_EPSR}.  

Fewer researchers have investigated the effects of planning mitigation methods for wildfires. 
Those studies have included planning mitigation methods that typically focus on long-term decisions such as hardening investments, Distributed Energy Resources (DER) placement, and risk-based system design \cite{Pianco_2025_IEEE_TPS,Ma_2018_IEEETSG, Yuan_2016_IEEETSG, Sohrabi_2024_IEEEA, Vahedi_2025_}.
While some studies integrate both planning and operations by combining hardening, PSPS, DERs, and microgrid formation within one mitigation analysis \cite{Yao_2025_IJEPES, Ganguly_2025_EJOR}, these studies do not explore the impact of these mitigation actions across both Tx~\&~Dx systems and therefore do not capture the voltage interdependence or cascading wildfire-induced interactions between Tx~\&~Dx-level resilience strategies.
In addition to mitigation strategy, another key dimension across the literature is the choice of electrical system type, which determines whether wildfire impacts are analyzed at the Tx or Dx level \cite{Nazaripouya_2020_IEEEPESGM}. 
% %%%%%%%%%%%%%%%%%%%%%%%%%%%%%%%%%
% Table Part 1
% %%%%%%%%%%%%%%%%%%%%%%%%%%%%%%%%%%
\FloatBarrier
\afterpage{%
\begin{sidewaystable*}[p]
\caption{Literature on Electrical Resilience, Planning, and Assessment for Wildfires and Other Hazards}
\label{tab:lit_rotated}
\renewcommand{\arraystretch}{1.5}
\centering
\small
\begin{tabular}{
L{4cm} % Planning / Op
p{.3cm} %  Dx
p{.3cm} % Tx
L{2.2cm} % IEEE Case Study
L{1.5cm} % Hazard
L{1.5cm} % Component Exposure
L{2.2cm} % Component Response Modeling
L{.3cm} % Damage
L{.3cm} % Restoration
L{.3cm} % Recovery
L{2cm} % author
} % Social Component
\toprule
\textbf{Planning / Operations} &
\multicolumn{2}{c}{\textbf{System Type}} &
\textbf{System Case Study} &
\textbf{Hazard} &
\textbf{Component Exposure} &
\textbf{Component Response Modeling} &
\multicolumn{3}{c}{\textbf{Disaster Phase}} &
\textbf{Author (Year)} 
\\

& 
\multicolumn{1}{c}{\textbf{Tx}} &
\multicolumn{1}{c}{\textbf{Dx}} &
& & & &
\multicolumn{1}{c}{\textbf{DMG}} &
\multicolumn{1}{c}{\textbf{RES}} &
\multicolumn{1}{c}{\textbf{REC}} &
\\
\midrule

Operations (pre-disaster and during-event dispatch of resources) &
\multicolumn{1}{c}{$\square$} &
\multicolumn{1}{c}{$\blacksquare$} &
Modified IEEE 123-bus &
Wildfire & 
Lines & 
Performance degradation model &
\multicolumn{1}{c}{$\blacksquare$} &
\multicolumn{1}{c}{$\square$} &
\multicolumn{1}{c}{$\square$} &
Ansari \& Mohagheghi (2014) \cite{Ansari_2014_ITEES}
\\

Both (OPF and preignition mitigation) &
\multicolumn{1}{c}{$\blacksquare$} &
\multicolumn{1}{c}{$\square$} &
IEEE 30-bus &
Wildfire & 
--- & 
--- &
\multicolumn{1}{c}{$\blacksquare$} &
\multicolumn{1}{c}{$\square$} &
\multicolumn{1}{c}{$\square$} &
Sayarshad (2023) \cite{Sayarshad_2023_IJEPES}
\\

Operations (resilient scheduling under uncertainty) &
\multicolumn{1}{c}{$\square$} &
\multicolumn{1}{c}{$\blacksquare$} &
Modified 33-bus &
Wildfire & 
Lines &
Deterministic Heat-transfer &
\multicolumn{1}{c}{$\blacksquare$} &
\multicolumn{1}{c}{$\square$} &
\multicolumn{1}{c}{$\square$} &
Izadi et al. (2023) \cite{Izadi_2023_AE}
\\

Operations (capacity adjustment and vulnerability analysis) &
\multicolumn{1}{c}{$\blacksquare$} &
\multicolumn{1}{c}{$\square$} &
IEEE 30-bus &
Wildfire & 
Lines &  
Deterministic Heat-transfer &
\multicolumn{1}{c}{$\blacksquare$} &
\multicolumn{1}{c}{$\square$} &
\multicolumn{1}{c}{$\square$} &
Choobineh et al. (2015) \cite{Choobineh_2015_FSJ}
\\

Operations (post-disaster microgrid formation) &
\multicolumn{1}{c}{$\square$} &
\multicolumn{1}{c}{$\blacksquare$} &
IEEE 37-bus &
--- & --- & --- &
\multicolumn{1}{c}{$\blacksquare$} &
\multicolumn{1}{c}{$\blacksquare$} &
\multicolumn{1}{c}{$\square$} &
Zhong et al. (2024) \cite{Zhong_2024_EL}
\\

Operations (power restoration w/ emergency comm vehicles) &
\multicolumn{1}{c}{$\square$} &
\multicolumn{1}{c}{$\blacksquare$} &
IEEE 123-bus &
--- & --- & --- &
\multicolumn{1}{c}{$\blacksquare$} &
\multicolumn{1}{c}{$\blacksquare$} &
\multicolumn{1}{c}{$\square$} &
Ye et al. (2023) \cite{Ye_2023_IEEESG}
\\

Planning (hardening investments) &
\multicolumn{1}{c}{$\square$} &
\multicolumn{1}{c}{$\blacksquare$} &
Modified EPRI 74-mile circuit &
Hurricanes &  
Lines &  
Component fragility &
\multicolumn{1}{c}{$\blacksquare$} &
\multicolumn{1}{c}{$\blacksquare$} &
\multicolumn{1}{c}{$\square$} &
Ma et al. (2018) \cite{Ma_2018_IEEETSG}
\\

Operations (microgrid-based load restoration) &
\multicolumn{1}{c}{$\square$} &
\multicolumn{1}{c}{$\blacksquare$} &
IEEE 37-bus and 123-bus &
--- & --- & --- &
\multicolumn{1}{c}{$\blacksquare$} &
\multicolumn{1}{c}{$\blacksquare$} &
\multicolumn{1}{c}{$\square$} &
Chen et al. (2016) \cite{Chen_2016_TSG}
\\

Planning (optimization for hardening and DG placement) &
\multicolumn{1}{c}{$\square$} &
\multicolumn{1}{c}{$\blacksquare$} &
IEEE 33-bus and 123-bus &
Natural Hazards &  
Lines &  
Binary failure  &
\multicolumn{1}{c}{$\blacksquare$} &
\multicolumn{1}{c}{$\square$} &
\multicolumn{1}{c}{$\square$} &
Yuan et al. (2016) \cite{Yuan_2016_IEEETSG}
\\

Both (grid resilience review paper for wildfires) &
\multicolumn{1}{c}{$\square$} &
\multicolumn{1}{c}{$\square$} &
--- &
Wildfire & 
Grid & 
Multiple &
%Both (review of wildfire damage and restoration) &
\multicolumn{1}{c}{$\blacksquare$} &
\multicolumn{1}{c}{$\blacksquare$} &
\multicolumn{1}{c}{$\square$} &
Nazaripouya (2020) \cite{Nazaripouya_2020_IEEEPESGM}
\\

Operations (dynamic line rating and resource management) &
\multicolumn{1}{c}{$\square$} &
\multicolumn{1}{c}{$\blacksquare$} &
Modified IEEE 33-bus &
Wildfire & 
Lines &
Deterministic Heat-transfer &
%Damage modeling (reduced line capacity; no restoration &
\multicolumn{1}{c}{$\blacksquare$} &
\multicolumn{1}{c}{$\square$} &
\multicolumn{1}{c}{$\square$} &
Trakas and Hatziargyriou (2018) \cite{Trakas_2018_IEEETPS}
\\ 

Operations (wildfire risk-aware operation planning) &
\multicolumn{1}{c}{$\blacksquare$} &
\multicolumn{1}{c}{$\square$} &
6-bus and IEEE 30-bus &
Wind &
Lines &
Physics-based conductor clashing model &
%--- &
\multicolumn{1}{c}{$\blacksquare$} &
\multicolumn{1}{c}{$\square$} &
\multicolumn{1}{c}{$\square$} &
Bayani et al. (2023) \cite{Bayani_2023_IEEESJ}
\\

Planning (new lines, line hardening, switch installation) &
\multicolumn{1}{c}{$\square$} &
\multicolumn{1}{c}{$\blacksquare$} &
54-bus (see \cite{Moreira_2024_IEEE_TPS})&
Wildfire &
Lines &
Binary availability &
%--- &
\multicolumn{1}{c}{$\blacksquare$} &
\multicolumn{1}{c}{$\square$} &
\multicolumn{1}{c}{$\square$} &
Pianc\'{o} et al. (2025) \cite{Pianco_2025_IEEE_TPS}
\\

\bottomrule
\end{tabular}
\end{sidewaystable*}
% %%%%%%%%%%%%%%%%%%%%%%%%%%%%%%%%%%
% Table Part 2
% %%%%%%%%%%%%%%%%%%%%%%%%%%%%%%%%%%
\begin{sidewaystable*}[p]
\caption{Continued}
\label{tab:lit_rotated2}
\renewcommand{\arraystretch}{1.5}
\centering
\small
\begin{tabular}{
L{4cm} % Planning / Op
p{.3cm} %  Dx
p{.3cm} % Tx
L{2.2cm} % IEEE Case Study
L{1.5cm} % Hazard
L{1.5cm} % Component Exposure
L{2.2cm} % Component Response Modeling
L{.3cm} % Damage
L{.3cm} % Restoration
L{.3cm} % Recovery
L{2cm} % author
} % Social Component
\toprule
\textbf{Planning / Operations} &
\multicolumn{2}{c}{\textbf{System Type}} &
\textbf{System Case Study} &
\textbf{Hazard} &
\textbf{Component Exposure} &
\textbf{Component Response Modeling} &
\multicolumn{3}{c}{\textbf{Disaster Phase}} &
\textbf{Author (Year)} 
\\
& 
\multicolumn{1}{c}{\textbf{Tx}} &
\multicolumn{1}{c}{\textbf{Dx}} &
& & & &
\multicolumn{1}{c}{\textbf{DMG}} &
\multicolumn{1}{c}{\textbf{RES}} &
\multicolumn{1}{c}{\textbf{REC}} &
\\
\midrule

Operations (real-time operation during wildfire) &
\multicolumn{1}{c}{$\square$} &
\multicolumn{1}{c}{$\blacksquare$} &
Modified IEEE 123-bus &
Wildfire & 
Lines &  
Deterministic Heat-transfer &
\multicolumn{1}{c}{$\blacksquare$} &
\multicolumn{1}{c}{$\square$} &
\multicolumn{1}{c}{$\square$} &
Mohagheghi \& Rebennack (2015) \cite{Mohagheghi_2015_IJEPES}
\\

Operations (attacker–defender wildfire resilience) &
\multicolumn{1}{c}{$\blacksquare$} &
\multicolumn{1}{c}{$\square$} &
Chilean 65-bus and 278-bus &
Wildfire &  
Lines & 
Binary failure &
\multicolumn{1}{c}{$\blacksquare$} &
\multicolumn{1}{c}{$\square$} &
\multicolumn{1}{c}{$\square$} &
Tapia et al. (2021) \cite{Tapia_2021_EJOR}
\\

Operations (PSPS optimization) &
\multicolumn{1}{c}{$\blacksquare$} &
\multicolumn{1}{c}{$\square$} &
RTS-GMLC &
Wildfire &  
Lines &  
Binary availability &
\multicolumn{1}{c}{$\square$} &
\multicolumn{1}{c}{$\square$} &
\multicolumn{1}{c}{$\square$} &
Rhodes et al. (2021) \cite{Rhodes_2021_IEEETPS}
\\

Both (hardening, DG, PSPS and microgrid formation) &
\multicolumn{1}{c}{$\square$} &
\multicolumn{1}{c}{$\blacksquare$} &
IEEE 33-bus &
Wildfire &
Lines &
Binary failure &
\multicolumn{1}{c}{$\blacksquare$} &
\multicolumn{1}{c}{$\blacksquare$} &
\multicolumn{1}{c}{$\square$} &
Yao et al. (2025) \cite{Yao_2025_IJEPES}
\\

Both (hardening and microgrid recovery) &
\multicolumn{1}{c}{$\square$} &
\multicolumn{1}{c}{$\blacksquare$} &
IEEE 14-bus and 30-bus &
Wildfire &  
Buses and lines &  
Binary failure of components &
\multicolumn{1}{c}{$\blacksquare$} &
\multicolumn{1}{c}{$\blacksquare$} &
\multicolumn{1}{c}{$\square$} &
Ganguly et al. (2025) \cite{Ganguly_2025_EJOR}
\\

Planning (AC power flow and risk) &
\multicolumn{1}{c}{$\blacksquare$} &
\multicolumn{1}{c}{$\square$} &
IEEE 30-bus &
Wildfire &
Lines &
Binary failure &
\multicolumn{1}{c}{$\blacksquare$} &
\multicolumn{1}{c}{$\square$} &
\multicolumn{1}{c}{$\square$} &
Sohrabi et al. (2024) \cite{Sohrabi_2024_IEEEA}
\\

Operations (PSPS and DG dispatch) &
\multicolumn{1}{c}{$\blacksquare$} &
\multicolumn{1}{c}{$\square$} &
IEEE 14-bus and 24-bus &
Wildfire &
Lines &
Binary availability &
%--- &
\multicolumn{1}{c}{$\square$} &
\multicolumn{1}{c}{$\square$} &
\multicolumn{1}{c}{$\square$} &
Greenough et al. (2025) \cite{Greenough_2025_IEEETPS}
\\

Operations (OPF and generator dispatch) &
\multicolumn{1}{c}{$\blacksquare$} &
\multicolumn{1}{c}{$\square$} &
IEEE 14-bus &
Wildfire &
Lines &
Deterministic Heat-transfer &
\multicolumn{1}{c}{$\blacksquare$} &
\multicolumn{1}{c}{$\square$} &
\multicolumn{1}{c}{$\square$} &
Nematshahi et al. (2026) \cite{Nematshahi_2026_EPSR}
\\

Operations (real-time, self-proactive control) &
\multicolumn{1}{c}{$\square$} &
\multicolumn{1}{c}{$\blacksquare$} &
Modified IEEE-38 and IEEE 123-bus &
Wildfire &
Lines &
Deterministic Heat-transfer &
\multicolumn{1}{c}{$\blacksquare$} &
\multicolumn{1}{c}{$\blacksquare$} &
\multicolumn{1}{c}{$\square$} &
Shalaby et al. (2025) \cite{Shalaby_2025_,Shalaby_2025_IEEE_TPS}
\\

Planning (risk assessment) &
\multicolumn{1}{c}{$\blacksquare$} &
\multicolumn{1}{c}{$\square$} &
RTS-GMLC &
Wildfire &
Lines &
Component fragility &
\multicolumn{1}{c}{$\blacksquare$} &
\multicolumn{1}{c}{$\square$} &
\multicolumn{1}{c}{$\square$} &
Vahedi et al. (2026) \cite{Vahedi_2026_IEEE_TPS}
\\

Planning (hazard and exposure) &
\multicolumn{1}{c}{$\square$} &
\multicolumn{1}{c}{$\blacksquare$} &
Real feeder &
Wildfire &
Pole &
Component fragility &
\multicolumn{1}{c}{$\blacksquare$} &
\multicolumn{1}{c}{$\square$} &
\multicolumn{1}{c}{$\square$} &
Campos et al. (2025) \cite{Campos_2026_IJDRR}
\\

Operations (line de-energization, network reconfiguration, switching) &
\multicolumn{1}{c}{$\square$} &
\multicolumn{1}{c}{$\blacksquare$} &
54-bus and 138-bus &
Wildfire &
Lines &
Binary availability &
%--- &
\multicolumn{1}{c}{$\blacksquare$} &
\multicolumn{1}{c}{$\square$} &
\multicolumn{1}{c}{$\square$} &
Moreira et al. (2024) \cite{Moreira_2024_IEEE_TPS}
\\

Operations (PSPS and mobile power sources (MPS)) &
\multicolumn{1}{c}{$\square$} &
\multicolumn{1}{c}{$\blacksquare$} &
33-bus &
Wildfire &
Lines &
Binary availability &
\multicolumn{1}{c}{$\square$} &
\multicolumn{1}{c}{$\blacksquare$} &
\multicolumn{1}{c}{$\square$} &
Su et al. (2024) \cite{Su_2024_IEEE_TPS}
\\

\bottomrule
\end{tabular}
\end{sidewaystable*}
}

\subsection{System Type}
Wildfire-related electrical studies commonly focus on either Tx or Dx systems, where coupled Tx\&Dx studies are rare \cite{Li_2018_IEEETPS, Sun_2023_PESPSTP, Veeramany_2018_IEEESJ, Tang_2020_JPES, Nawaz_2021_JMPSCE, Wang_2022_IEEETSG, Selvaraj_2023_IETGTD, Lu_2025_EPSR} and no studies couple both systems in a wildfire context.
System type selection is decided based on the size of the study area, operational strategies, planning strategies,  and voltage-level focus.
Tx test systems do not consider downstream feeder behavior, which neglects community-level restoration and how Tx operation strategies affect restoration time.
These Tx\&Dx frameworks focus on non-hazard-related contingencies, where a component disconnection occurs independent of the physical hazard processes. 
Consequently, these studies do not consider the impact of wildfire propagation, self-induced ignition risk, or the downstream effects to communities, including disruptions to critical facilities.

A wildfire-aware test system should include a connected transmission and distribution network subjected to a realistic wildfire simulation that implements WUI and community components.
This type of integration allows for a more realistic assessment of resilience, risk, and the downstream effects to communities. 
A coupled Tx\&Dx test system could simulate the interdependencies of the two systems and their impact on community resilience.
Once the system type is defined, the selected modeling approach is implemented using standardized test systems that provide a reproducible benchmark for evaluating mitigation and resilience strategies.

\subsection{System Case Study}\label{sec: SystCaseStud}
IEEE standard test systems are useful for comparing methods across various studies because they are reproducible system benchmarks. 
In addition, real-world systems are often not used in research due to security concerns and data availability \cite{Tapia_2021_EJOR, Ma_2018_IEEETSG, Campos_2026_IJDRR}. 
A standard test system incorporates several key components that are configured to represent theoretical electrical behavior, including buses, lines, generation, and load.
Here, buses represent nodes in the network where the voltage is defined, lines connect buses, transformers connect buses at different voltages, generation supplies power, and loads represent the power demand.
To research Tx level components and systems, analyses often utilize IEEE 14-bus, 30-bus, or RTS-GMLC test systems to study dispatch, de-energization, and risk \cite{Sayarshad_2023_IJEPES, Rhodes_2021_IEEETPS, Sohrabi_2024_IEEEA, Vahedi_2025_, Bayani_2023_IEEESJ}.
To research Dx level components and systems, analyses use IEEE 33-bus, 37-node, and 123-node feeders to study feeder outages, operational mitigation, and local restoration \cite{Izadi_2023_AE, Chen_2016_TSG, Ye_2023_IEEESG, Zhong_2024_EL, Yao_2025_IJEPES}.
Additionally, Dx studies use IEEE feeders with modifications (e.g., normally open tie switch, switching logic, and assumed DER availability) \cite{Shalaby_2025_IEEE_TPS,Trakas_2018_IEEETPS,Chen_2016_TSG, Ye_2023_IEEESG, Yao_2025_IJEPES}.

To explore specific mitigation and operational strategies during a wildfire, researchers often modify standard test systems, creating new systems along with unique assumptions that limit comparability and make it difficult to extend prior research efforts.
The RTS-GMLC \cite{RTS_GMLC} is one example of a test system that attempts to address this problem by standardizing commonly used modifications and incorporating them directly into the test system, which allows researchers to vary operational methods and decision strategies without modifying the system.
This system is primarily implemented to study Tx level operational and planning mitigation strategies \cite{Rhodes_2021_IEEETPS,Vahedi_2026_IEEE_TPS}.  
However, wildfire-electrical resilience studies currently lacks an analogous case system, where current studies use a wide range of assumptions for wildfire modeling, fuel landscapes, system level (Tx or Dx), exposure modeling, and component response modeling. 
The result is that researchers are using different types of systems, modifying systems, and imposing various planning or operational tactics to different test systems, making it difficult to compare the results between research studies. 
Therefore, it is challenging to draw broader conclusions about most effective mitigation and operational methods for wildfire-resilient electrical infrastructure from the broader body of knowledge.
While standardized case test systems define the electrical system structure, the wildfire resilience analyses will ultimately depend on how well the fire hazard itself is represented within these systems.

\subsection{Hazard}
Fire modeling in power systems research studies can generally be placed into three categories \cite{Choobineh_2015_FSJ}:~theoretical \cite{Mohagheghi_2015_IJEPES, Choobineh_2015_FSJ, Izadi_2023_AE, Trakas_2018_IEEETPS} (based on first principles or fire physics), empirical \cite{Shalaby_2025_IEEE_TPS, Yao_2025_IJEPES, Greenough_2025_IEEETPS, Su_2024_IEEE_TPS, Moreira_2024_IEEE_TPS, Rhodes_2021_IEEETPS} (relying on statistical or probability correlations), or semi-empirical \cite{Ganguly_2025_EJOR, Tapia_2021_EJOR, Sayarshad_2023_IJEPES, Ansari_2014_ITEES, Nematshahi_2026_EPSR, Shalaby_2025_, Pianco_2025_IEEE_TPS, Vahedi_2026_IEEE_TPS, Campos_2026_IJDRR} (combines simplified physical equations and experimental data).
Each method has its limitations where the type of fire modeling is chosen based on the scope of the study, required exposure characteristics (spatial versus accurate intensity measurements), and fidelity of the study.
For example, a large study area may benefit from computationally efficient fire spread simulations using semi-empirical fire models (e.g., FARSITE \cite{Campos_2026_IJDRR, Nematshahi_2026_EPSR}, FlamMap \cite{Ganguly_2025_EJOR}, or Rothermel's equations \cite{Sayarshad_2023_IJEPES}) as opposed to a simulation with high computational cost such as a theoretical model that implements computational fluid mechanics.

Wildfire ignitions may be caused by lightning or by human activity, with human-caused ignitions accounting for the majority of reported wildfire events \cite{Radeloff, Campos_2026_IJDRR,Bayani_2023_IEEESJ}. 
Under extreme fire weather conditions, self-induced electrical ignitions may occur due to line faults, tower failures, or pole failures, where the subsequent fire spread is driven by high wind conditions that propagate the fire through the landscape via convection and firebrands or embers.
Recent studies have begun to consider these electrical self-induced ignitions and their implications for power system performance \cite{Moreira_2024_IEEE_TPS,Rhodes_2021_IEEETPS,Su_2024_IEEE_TPS,Pianco_2025_IEEE_TPS,Greenough_2025_IEEETPS,Ganguly_2025_EJOR,Yao_2025_IJEPES,Sayarshad_2023_IJEPES, Bayani_2023_IEEESJ, Campos_2026_IJDRR}. 
However, few studies explore the power outages that may result from the behavior of electrical systems under high wind conditions (e.g., conductor clashes, vegetation strikes, and conductor splice failures). 
After simulating the ignitions due to electrical components, the simulation of the fire propagation itself is critical to quantify the spatial distribution of the potential fire as well as the exposure of grid components. 
Few studies model fire propagation after a self-induced ignition, where the fire is typically modeled using some form of Rothermel's equations for fire propagation \cite{Ganguly_2025_EJOR,Sayarshad_2023_IJEPES, Campos_2026_IJDRR}. 

Realistic hazard simulation is a foundational component of risk and resilience studies.
Resilience studies often include operational decisions being made temporally throughout the duration of the fire. 
Therefore, if a wildfire simulation does not propagate in a natural manner, then the results may be misleading, and if implemented, could lead to mitigation strategies that fail to meet community resilience objectives.
Once the wildfire hazard methodology has been defined, the next critical consideration is how that hazard translates into exposure of specific electrical components.

\subsection{Component Exposure}
Overhead lines are the dominant exposed component class for both Tx~\&~Dx within previous research \cite{Ansari_2014_ITEES, Izadi_2023_AE, Choobineh_2015_FSJ,Yuan_2016_IEEETSG,Trakas_2018_IEEETPS,Mohagheghi_2015_IJEPES, Tapia_2021_EJOR,Rhodes_2021_IEEETPS,Yao_2025_IJEPES, Ganguly_2025_EJOR,Sohrabi_2024_IEEEA,Nematshahi_2026_EPSR,Shalaby_2025_,Vahedi_2025_}.  
This is primarily due to the ignition risk associated with conductors and the susceptibility of line capacity and protection to wildfire heat \cite{Rhodes_2021_IEEETPS, Trakas_2018_IEEETPS}.
Additionally, conductors span large distances, which increases the wildfire exposure probability.
A single study considers both buses and lines when modeling wildfire mitigation strategies for Dx \cite{Ganguly_2025_EJOR}.
Only limited work isolates poles as a distinct exposed asset class, despite their importance in distribution networks and their fire and wind vulnerability \cite{Campos_2026_IJDRR}.
Component exposure alone is insufficient to assess system impacts.
Therefore, researchers couple component exposure with explicit models that translate wildfire hazard exposure into electrical component-level damage or failure.

\subsection{Component Response Modeling}\label{Sec: CompRespModel}
Component response to wildfire exposure is commonly represented using binary failure models, deterministic derating models, or fragility-based probabilistic models.  
Binary failure models treat exposed components as either failed or intact (e.g., once a fire reaches a line, that line is considered to be in a failed state), which is useful for optimization-driven interdiction, PSPS, and risk studies \cite{Tapia_2021_EJOR, Rhodes_2021_IEEETPS, Sohrabi_2024_IEEEA, Vahedi_2025_}.
Comparatively, deterministic heat-transfer models evaluate whether conductor temperature exceeds operational limits, resulting in temporary line derating or disconnection \cite{Choobineh_2015_FSJ, Mohagheghi_2015_IJEPES, Trakas_2018_IEEETPS, Nematshahi_2026_EPSR, Shalaby_2025_}.
Fragility-based approaches model the probability of failure of electrical components given a fire intensity  \cite{Nazemi_2024_IEEE_TPS,Ma_2018_IEEETSG, Nazaripouya_2020_IEEEPESGM, Campos_2026_IJDRR, Vahedi_2026_IEEE_TPS}. 

The accuracy of modeling the component response depends on the availability of experimental testing data or data from real fires. When integrated within modeling frameworks, the ability of the model to accurately predict component responses will depend upon the validity of the hazard model itself. 
However, often simulation data, empirical data, or fragility functions are not available. 
In these cases, assumptions are made within the research. 
For example, binary response modeling does not require hazard intensity inputs. 
The component is considered failed if exposed to a fire with no regard for the intensity of the fire. 
In this case, the component response is highly dependent upon the fire perimeters of the hazard model. 
Whereas fragility models are functions of hazard intensity and therefore require realistic hazard modeling \cite{Campos_2026_IJDRR,Izadi_2023_AE,Tapia_2021_EJOR}.
These component-level response models form the foundation for modeling electrical system damage, restoration, and recovery.

\subsection{Disaster Phase:~Damage, Restoration, and Recovery}
\begin{figure}
    \centering
    \includegraphics[width=1\linewidth]{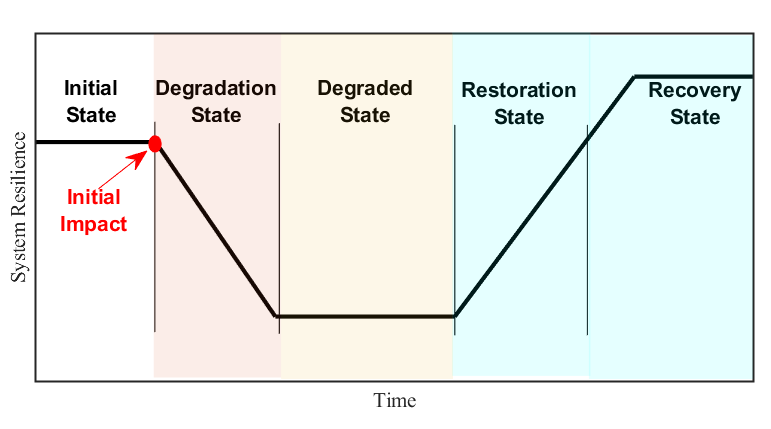}
    \caption{Visualization of a trapezoidal resilience curve and resilience states of an electrical system during a hazard event.}
    \label{fig:res_elec}
\end{figure}

Studies investigating wildfire impacts on electrical systems generally model the damage to the system through component degradation/failure (see Section~\ref{Sec: CompRespModel}) or probabilistic/scenario-based power disruptions.
The restoration and recovery of the system and the components can inform utilities about the resilience of their system as well as be a metric for social and economic recovery for the community. 
Restoration is often captured through operational-level action, such as, fast switching, microgrid formation, and self-healing network designs, which allows for re-energization post-hazard \cite{Chen_2016_TSG,Zhong_2024_EL}. 
A few Dx level studies consider the restoration process through microgrid islanding, reconfiguration, or crew coordination \cite{Chen_2016_TSG, Ye_2023_IEEESG, Zhong_2024_EL}.
System recovery may be considered to be a long-term process, whereby a system is repaired and restored to pre-event functionality or better (Figure \ref{fig:res_elec}) \cite{NIST_CRPGBIS_2020}. 
Long-term recovery is achieved through repair crew dispatch to restore damaged components such as poles and lines, returning the system to full functionality while operational strategies maintain a stable operating state during restoration.
Figure~\ref{fig:res_elec} shows a typical resilience curve and outlines the progression of an electrical system from the initial impact to fully recovered, consistent with the four Rs of resilience: robustness, redundancy, resourcefulness, and rapidity.
Literature shows that studies tend to quantify performance impacts after outages or reduced line capacity, where restoration efforts are limited and recovery modeling is absent \cite{Ansari_2014_ITEES, Choobineh_2015_FSJ, Sayarshad_2023_IJEPES, Trakas_2018_IEEETPS, Mohagheghi_2015_IJEPES}.  
Damage, restoration, and recovery describe system-level performance; however, their significance is realized through the social and economic impacts experienced by the communities served by the electrical system.

\subsection{Community Component}\label{sec: CommComp}
The impact of wildfires on communities, compounded by power outages, can have significant social and economic impacts to a community \cite{Sakellariou_2022_IJDRR}. 
Power outages during hot weather conditions can leave residents without access to refrigerated food, health care processes that require electricity, or climate control mechanisms within their homes. 
PSPS can force business closures within communities and leave critical infrastructure and facilities inoperable. 
The self-induced fire impact and resulting damage to critical community assets is devastating due to the potentially lengthy restoration process of fire-damaged buildings; therefore, studies should consider the social and economic aspects of wildfire risk that electrical systems impose on the communities they serve.
This is a consistent limitation across wildfire-electrical resilience studies: the absence of explicit social and economic impacts to communities.

The identified gaps in the literature and existing test-system limitations (see Sections~\ref{sec: OperPlann} through \ref{sec: CommComp}) have important implications for wildfire resilience research. 
Current studies often omit critical components needed to accurately assess community-level resilience, a shortcoming that is further exacerbated by the absence of a standardized wildfire-focused test system. 
Therefore, to address these limitations, a framework that can supplement such a test system is necessary. 
Accordingly, the following section proposes a framework designed to advance wildfire-electrical resilience studies (Figure~\ref{fig:Framework}).

\section{Proposed Framework}\label{sec: Proposed_Framework}
\begin{figure*}[t!]
    \centering
    \includegraphics[width=1.0\textwidth]{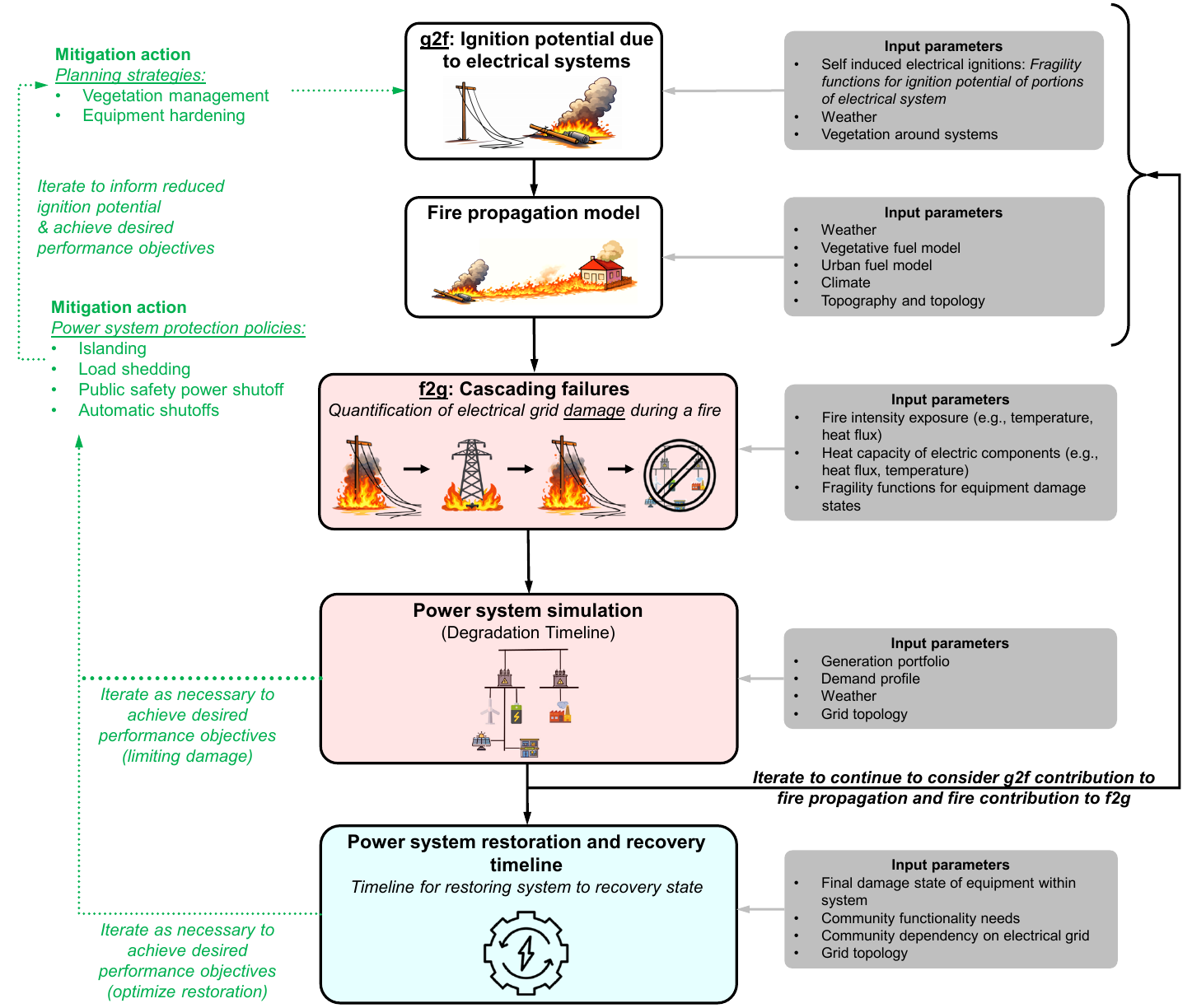}
        \caption{Framework implementing key requirements for a wildfire-aware test system.}
    \label{fig:Framework}
\end{figure*}

In this Section we detail the proposed framework for identifying and jointly modeling electrical system vulnerabilities and wildfire risk. 
Figure~\ref{fig:Framework} summarizes the proposed framework and shows how it incorporates the key dimensions from Table~\ref{tab:lit_rotated} and research gaps identified in Section~\ref{sec: Background}.
The framework implements a typical resilience/performance curve structure, where the scenario begins with an initial system impact followed by degradation, restoration, and recovery (see Fig.~\ref{fig:res_elec}).
Additionally, it integrates a coupled Tx~\&~Dx network within a WUI landscape, where wildfire hazard inputs interact dynamically with electrical components.
Operational mitigation and planning resilience actions are simulated alongside social and community inputs or metrics.

The framework encompasses two fundamental interacting scenarios, g2f and f2g.
These bidirectional electrical–wildfire interactions are imperative for electrical system resilience assessments because they allow wildfire-electrical grid simulations to move beyond static, one-dimensional damage representations and instead simulate the interactions between electrical infrastructure and wildfire hazard.
Implementing both g2f and f2g allows for the explicit modeling of electrical self-induced ignitions in conjunction with wildfire-induced electrical damage.
Thus, the framework reflects the underlying nature of electrical systems as both the potential source and recipient of ignitions, fires, and cascading disruptions. 
Given that wind is the driver for electrical infrastructure failure during fire weather conditions, the bidirectional coupling is compounded, as wind contributes to g2f through wind-induced component failure and also exacerbates f2g through faster fire propagation, leading to increased component exposure. 
Ultimately, these integrated bidirectional interactions allow researchers to better quantify cascading failures, as well as evaluate mitigation strategies which are key contributions to overall system risk and wide-area resilience studies. 

\subsection{Grid-to-fire (g2f) scenario}
The g2f scenario considers the ignition potential of surrounding fuel sources caused by the electrical system due to wind-induced electrical component failures, faults from conductor clashes, or vegetation clashes from strong winds (i.e., initial impact).
Under fire weather conditions, strong winds are responsible for the initial electrical component failure, triggering an ignition; however, after the ignition, direct fire exposure becomes the driving hazard for subsequent electrical infrastructure damage.

The propagation of a fire due to a self-induced ignition can be simulated with a fire behavior simulator, such as WRF-Fire\cite{Skamarock_2019_NCAR} or FARSITE \cite{missoula_fire_lab_cmd_apps}, to produce fire perimeters based on local parameters such as weather (wind, humidity, precipitation, and temperature), vegetation and urban fuel models, climate, and topography/topology.
The fire modeling may be simulated on time scales (e.g., 30 minute timestep) that are consistent with real world operational actions time frames.
Planning-level mitigation actions, including vegetation management and equipment hardening, are incorporated at the g2f ignition potential evaluation stage for the model to account for the reduced probability of self-induced ignitions and in turn, reduce the frequency of f2g occurrence.
Consequently, reducing the initial wind-driven ignition through 
these actions lowers fire risk by interrupting the hazard transition from wind-driven failures to fire-driven cascading damage.
Further, network resilience is improved by increasing the network's robustness through planning actions (i.e., decreasing asset damage), which reduces restoration and recovery efforts.

If a self-induced ignition occurs, then operational-level mitigation actions (power system protection policies) are implemented to limit damage to local equipment, propagation of the failures, and potential power loss to customers, and critical infrastructure. 
These local power system protection policies can be further coordinated at the system level, to capture specific combinations of actions that alleviate a known wide-area vulnerability. 
They are commonly known as Remedial Action Schemes (RAS) or Special Protection Schemes (SPS).

The ability of the utility and grid to implement mitigation through power system protection policies, such as load shedding, islanding, public safety power shutoffs, will inform how much planning mitigation is required.
The reduction in fire risk is then achieved by changing the network configuration or service, which improves system robustness to fire-induced damage.
By modifying network configuration or de-energizing vulnerable segments, the system increases robustness to wildfire damage, relies on resourcefulness through active operational control, may temporarily reduce redundancy, and reduces the burden on rapidity by limiting post-event infrastructure damage and restoration demand.
However, the effectiveness of operational mitigation strategies depend on the network topology, topography, and the serviced community. 
The ability to implement these power system protection policies depends upon the grid that is being modeled as well as the effectiveness of planning mitigation actions. 
The explicit modeling of g2f is therefore critical in the evaluation in electrical network risk and resilience.
However, equally as important is the f2g scenario.

\subsection{Fire-to-grid (f2g) scenario}
The f2g considers fire impacts on the grid components.
At this stage in the framework, a damage assessment of the grid components is realized using the fire propagation model and physics-based heat transfer equations, heat capacity of components, or fragility functions.
Within the f2g modeling, fire exposure becomes the prominent damage driver and as a results, the damaged component defines the consequence term in risk to the electrical system.

Further, the effects of cascading failure must be incorporated, so we can track the progression between sets of initiating events (exogenous events) and the mechanisms by which they propagate further with subsequent dependent failures (endogenous events).
Unlike the wind-induced ignitions in g2f, f2g cascading scenarios are determined by fire-induced component damage, which may cause secondary network failures. 
As a result, fire induced damage to the network shows how vulnerable components translate to network-wide consequences.
Further, Tx-induced fires may reach Dx components (or vice versa) and cause network-wide outages.
Additionally, Tx power outages from damaged components or operational actions cascade down to the Dx level.
In contrast to g2f, robustness decreases due to damaged components, redundancy decreases as the network degrades, resourcefulness is required to manage cascading outages, and rapidity is stressed due to accumulating component damage and extended restoration demand.

\subsection{Power system simulation}
At each power simulation time step, fire spread outputs (e.g., flame length, temperature, heat flux, and perimeter location) are mapped to electrical components to determine exposure levels. 
Component damage states are then evaluated using thermal response models or fragility functions, and failed components are removed from the network model. 
The updated network model is subsequently used to solve power flow and system state equations as desired.
The damage assessment and cascading effects are evaluated at each time step, allowing feedback between fire progression and grid response throughout the simulation. 
For example, f2g effects may cause more g2f effects, thereby exacerbating the fire propagation. 
This interaction allows for a better representation of the fire dynamics and time-evolving wildfire risk and resilience performance quantification.
In addition, this modeling methodology enables varying mitigation actions to be evaluated to reach desired performance objectives by the community. 

\subsection{Power system restoration and recovery}
Restoration and recovery phases after the containment of the fire can reflect the effectiveness of the imposed mitigation actions. 
Short-term, restoration prioritizes the community's needs, guided by social and community-level metrics, to restore power to valued assets and basic dependencies. 
Modifications of the operational and planning mitigation strategies can be evaluated to achieve the desired performance objectives for the community.
The restoration and recovery phases will not be simultaneous for all portions of the community; consequently, the utility’s and grid’s RAS capabilities may be critical for phased restoration.
Therefore, mitigation actions must be evaluated not only on the ability to reduce g2f and f2g, but also to enable restoration and recovery post-fire.

\subsection{Necessities for a Wildfire Resilience Test System}
The developed framework consists of two interacting scenarios (g2f and f2g) that capture feedback between wildfire ignition and propagation, electrical system damage and cascading failures, mitigation actions, and power system restoration and recovery. 
The resulting effect is the ability to evaluate the wildfire risk and resilience of electrical systems and the communities they serve, informed by the synthesis of existing studies and the identified gaps in current wildfire-electrical modeling approaches (see Section \ref{sec: Background}).
The framework is intended to support a wildfire-aware test system that accounts for the following key dimensions: coupled Tx~\&~Dx modeling, natural and self-induced wildfires, WUI landscapes, operation and planning mitigation, restoration and recovery timeline, and community functionality and dependency.

\section{Conclusions}
This study develops and recommends a wildfire-aware modeling framework grounded in state-of-the-art electrical test system case studies and wildfire modeling methods. Standard IEEE transmission or distribution testbeds do not represent WUI fire propagation in space and time, wind-driven self-induced ignitions, coupled Tx–Dx interactions, or community-level outage consequences. As a result, current modeling approaches may underestimate ignition likelihood, cascading infrastructure damage, and the full impact of wildfire exposure on electrical systems and the communities they serve. In addition, researchers and electrical utilities are limited in their abilities to prioritize mitigation actions and locations that will have the greatest post-fire restoration impact to communities. Explicit integration of g2f and f2g processes within a coupled simulation environment provides a necessary foundation to begin to evaluate electrical grid wildfire resilience and risk assessments that can consider that electrical systems may both act as the ignition trigger for the fire as well as be damaged by the fire itself.

% ---------- References ----------
\bibliographystyle{IEEEtran}
\bibliography{IEEE.bib}
\balance
\begin{IEEEbiographynophoto}{Richard Campos}
received his M.S. and Ph.D from the University of Oklahoma in civil engineering with a focus on structural and resilience engineering. He is currently a Postdoctoral Scholar at Oregon State University. His research interests focuses on wildfire risk and resilience of electrical infrastructure and communities. He has worked with NSF EPSCoR projects to provide resilience strategies for infrastructure systems by analyzing hazards, risk, and fragility of electrical and transportation infrastructure components.
\end{IEEEbiographynophoto}
\begin{IEEEbiographynophoto}{Erica Fischer}
received a B.S. from Cornell University and a Ph.D. from Purdue University. She is currently the Glenn Willis Holcomb Professor in Structural Engineering at Oregon State University. Her research interests involve innovative approaches to the resilience and robustness of civil infrastructure systems. She has led NSF-sponsored teams in post wildfire investigations after the 2018 Camp Fire, 2021 Marshall Fire, and 2025 LA Fires.
\end{IEEEbiographynophoto}
\begin{IEEEbiographynophoto}{Eduardo Cotilla-Sanchez}
(Senior Member, IEEE) received the M.S. and Ph.D. degrees in electrical engineering from The University of Vermont, Burlington, VT, USA, in 2009 and 2012, respectively. He is currently a Professor and an Associate Dean with the College of Engineering, Oregon State University, Corvallis, OR, USA. His research interest includes electrical infrastructure resilience and protection, in particular, the study of cascading outages. He is the Vice Chair of the IEEE Working Group on Cascading Failures.
\end{IEEEbiographynophoto}

\end{document}